\documentclass[prl,twocolumn,superscriptaddress]{revtex4}
\usepackage{graphicx}
\usepackage{dcolumn}
\usepackage{amsmath,amsthm,amssymb}
\usepackage{subfigure}

\begin{document}
\title{Comment on "Holographic Thermalization, stability of AdS, and the Fermi-Pasta-Ulam-Tsingou
paradox" by V. Balasubramanian et al.}

\author{Piotr Bizo\'n}
\affiliation{Institute of Physics, Jagiellonian
University, Krak\'ow, Poland}
  \affiliation{Max-Planck-Institut f\"ur Gravitationsphysik,
Albert-Einstein-Institut, Golm, Germany}
\author{Andrzej Rostworowski}
\affiliation{Institute of Physics, Jagiellonian
University, Krak\'ow, Poland}
\date{\today}

\maketitle

A recent interesting paper \cite{v} revisits the problem of (in)stability of AdS under spherically symmetric massless scalar field perturbations, first studied in \cite{br}. The authors  claim that the set of initial data which do not
trigger  instability is  larger  than originally envisioned in \cite{br}, in particular it comprises
small amplitude two-mode initial data with energy equally distributed
among the modes.  To support this claim, the long time  numerical evolution of these data was shown to be stable against black hole formation. The aim of this comment is to demonstrate that this numerical result is incorrect. Hereafter, we use the notation and  references to equations of~\cite{v}.

Using our code (see \cite{mr2} for the detailed description), we solved  the   system of equations (2-4)  for the  two-mode initial data (20) with $\kappa=3/5$ and  $\varepsilon=0.09$ used in Fig.~3 in \cite{v}. The comparison of our result with the one of \cite{v} is shown in Fig.~1 which depicts the upper envelope of the quantity $\Pi^2(t,0)$ (related linearly to the Ricci scalar at the origin).
\begin{figure}[h!]
  \includegraphics[width=0.48\textwidth]{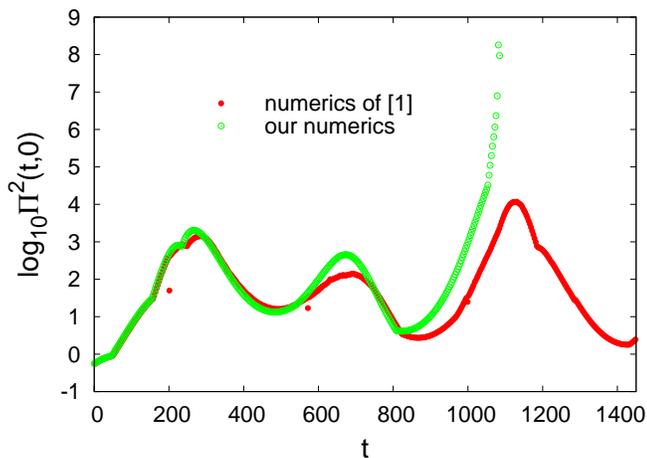}
   \caption{The upper envelope of $\Pi^2(t,0)$  for solutions starting from the two-mode initial data (20) with $\kappa=3/5$ and $\varepsilon=0.09$.
  Superimposed (red curve) is the numerical result of \cite{v} . }
  \label{fig1}
\end{figure}

Until the first local minimum at $t\approx 500$ the two curves stay together;  small discrepancies being due to different normalizations of time (our $t$ is the proper time at the origin while $t$ in \cite{v} is the proper time at infinity) and, probably, an inaccurate determination of the upper envelope of oscillations of $\Pi^2(t,0)$ in \cite{v}. However, for later times   the two curves begin to diverge. In particular, after the second local minimum we observe a rapid growth of the Ricci scalar at the origin and the formation of an apparent horizon  at $t\approx 1080$, whereas the numerical solution of \cite{v} remains bounded and enjoys a long (possibly infinite) lifetime.

 To  feel confident that our computation is correct (as opposed to the one of \cite{v}),
 we have validated it by  convergence tests.
  The evidence for the expected fourth-order convergence is shown in Fig.~2.
   \begin{figure} [h]
 \includegraphics[width=0.48\textwidth]{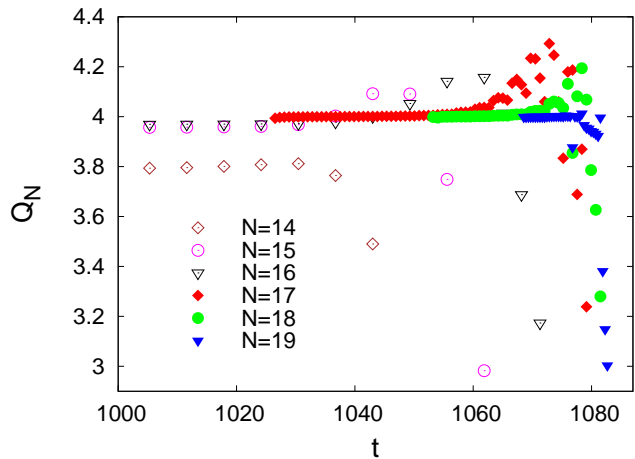}
  \caption{The late time convergence test for the simulation shown in Fig.~1 is depicted in blue.
  The convergence factor for the solution $\Phi_N$ computed on the grid of size $2^N$ is defined by
  $Q_N=\log_2\left(\frac{||\Phi_{N-2}-\Phi_{N-1}||}{||\Phi_{N-1}-\Phi_{N}||}\right)$, where $||\cdot||$ is the spatial $\ell_2$-norm. The convergence tests for coarser grids are added in order to illustrate the degradation of convergence caused by an insufficient  spatial resolution.
  }
 \label{fig2}
\end{figure}

We stress that in numerical simulations of turbulent phenomena the convergence test is an indispensable tool of verifying whether  small spatial scales are properly resolved.  We suspect that the numerical solution depicted by the red curve in Fig.~1 suffered from the gradual loss of spatial resolution (presumably due to a too coarse grid or/and ineffective adaptive mesh refinement) and, as a result of that, the simulation stepped over the  collapse and  went off track. Unfortunately, the `visual' convergence test shown in Fig.~3 of the Supplementary Material to \cite{v} was stopped much too early to spot the loss of resolution.

In conclusion, contrary to the claim made in \cite{v}, the question of
   existence of a threshold for black hole formation in the evolution of the two-mode  initial data (20) remains open. The resolution of this question seems very challenging  because the computational cost of simulations rapidly increases with $1/\varepsilon$.

\vskip 0.15cm \noindent \emph{Acknowledgments:} We thank the authors of \cite{v} for sending us the data file for the red curve in Fig.~1. This work was supported in part by the NCN grant NN202 030740.


\begin{thebibliography}{10}
\bibitem{v} V. Balasubramanian, A. Buchel, S.R. Green, L. Lehner, S.L. Liebling, \emph{
Holographic Thermalization, stability of AdS, and the Fermi-Pasta-Ulam-Tsingou paradox,} Phys. Rev. Lett. 113, 071601 (2014)

 \bibitem{br} P. Bizo\'n, A. Rostworowski, \emph{On weakly turbulent instability of anti-de Sitter space}, Phys. Rev. Lett. 107, 031102 (2011)

  \bibitem{mr2} M. Maliborski, A. Rostworowski, \emph{  Lecture Notes on Turbulent Instability of Anti-de Sitter Spacetime,}  International Journal of Modern Physics A, Vol. 28, 1340020 (2013)


\end{thebibliography}
\end{document}